# MANY-TO-MANY VOICE CONVERSION USING CONDITIONAL CYCLE-CONSISTENT ADVERSARIAL NETWORKS

*Shindong Lee[1,2], BongGu Ko[1], Keonnyeong Lee[1], In-Chul Yoo[1], and Dongsuk Yook[1]*

[1]Artificial Intelligence Laboratory, Department of Computer Science and Engineering, Korea University
[2]Multimedia R&D Part, AI Lab., Kakao Enterprise
koko9411@korea.ac.kr, {bgko, gnl0813, icyoo }@ai.korea.ac.kr, yook@korea.ac.kr

**ABSTRACT**

Voice conversion (VC) refers to transforming the speaker characteristics of an utterance without altering its linguistic contents. Many works on voice conversion require to have parallel training data that is highly expensive to acquire. Recently, the cycle-consistent adversarial network (CycleGAN), which does not require parallel training data, has been applied to voice conversion, showing the state-of-the-art performance. The CycleGAN based voice conversion, however, can be used only for a pair of speakers, i.e., one-to-one voice conversion between two speakers. In this paper, we extend the CycleGAN by conditioning the network on speakers. As a result, the proposed method can perform many-to-many voice conversion among multiple speakers using a single generative adversarial network (GAN). Compared to building multiple CycleGANs for each pair of speakers, the proposed method reduces the computational and spatial cost significantly without compromising the sound quality of the converted voice. Experimental results using the VCC2018 corpus confirm the efficiency of the proposed method.

## 1. INTRODUCTION

Voice conversion (VC) means modifying the speaker characteristics of a given speech while preserving the linguistic information. Many approaches for voice conversion have been studied that use parallel training data which contain pairs of the same transcription utterances spoken by different speakers [1–3]. Building such parallel data corpus is a highly expensive task, which has made strong needs for a voice conversion method that does not require the parallel training data.

Recently, several methods for voice conversion have been proposed that make use of non-parallel training data [4–11]. Among them, cycle-consistent adversarial network (CycleGAN) based approaches [10, 11] have shown the state-of-the-art sound quality of the converted speech without using any extra modules such as automatic speech recognition (ASR) systems. The CycleGAN [12], which is a variant of a generative adversarial network (GAN) [13], was originally developed to translate images between two different domains. By enforcing the cycle consistency using two GANs when applied to voice conversion, the CycleGAN can change the speaker related characteristics of an utterance without altering its linguistic contents. However, it learns only one-to-one mapping between two speakers. To achieve many-to-many voice conversion among $n$ speakers, $n(n-1)/2$ CycleGANs (i.e., $n(n-1)$ GANs) must be trained separately, which increases the training time and the memory space for model parameters prohibitively.

In this paper, we aim to extend the CycleGAN based approach [10, 11] to handle many-to-many voice conversion without too much computational burden during training while retaining the sound quality of the CycleGAN based voice conversion. In order to achieve this goal, we propose to use the conditional cycle-consistent adversarial network (CC-GAN) [14]. The GAN in a CC-GAN is conditioned on a pair of source and target speaker identity vectors, which is utilized to steer the GAN in synthesizing the speech for the target speaker. Since the single conditional GAN simulates the CycleGAN, it can make use of non-parallel training data as the CycleGAN but with less number of parameters. In this paper, we extend the CC-GAN to handle many-to-many voice conversion for more than two speakers. As the CC-GAN uses a single generator and discriminator to perform many-to-many voice conversion, the computational and spatial complexities are moderately managed while maintaining the voice quality of the CycleGAN based approach which is the state-of-the-art for one-to-one non-parallel voice conversion without using any extra modules.

The rest of the paper is organized as follows. Section 2 reviews the related works on voice conversion and describes the proposed method. Section 3 explains the experimental results, and Section 4 concludes the paper with some future research directions.

## 2. CONDITIONAL CYCLE-CONSISTENT ADVERSARIAL NETWORKS

### 2.1. Generative Adversarial Networks (GAN)

A GAN consists of two deep neural networks (DNN): a generator $G$ and a discriminator $D$ [13]. The generator is

---

trained to take a random noise $z$ as an input and generate an output that deceives the discriminator, while the discriminator is trained to classify whether the input is the real data $x$ or the fake ones made by the generator. The objective function of the GAN is defined as follows:

$$\mathcal{L}_{\text{GAN}}(G,D) = \mathbb{E}_x[\log D(x)] + \mathbb{E}_z[\log(1 - D(G(z)))] \, . \quad (1)$$

The generator is trained to minimize the objective function while the discriminator is trained to maximize it.

The vanilla GAN is not appropriate for voice conversion using non-parallel training data since there is no constraint that enforces to preserve the linguistic information of the input speech during conversion. The CycleGAN solves this problem by using a constraint called the cycle consistency loss.

## 2.2. Cycle-Consistent Adversarial Networks (CycleGAN)

The CycleGAN utilizes a pair of GANs, i.e., two generators ($G_X$ and $G_Y$) and two discriminators ($D_X$ and $D_Y$), where each GAN takes charge of one conversion direction. That is, generator $G_X$ converts the speech data $y$ from speaker $Y$ to that of speaker $X$, while generator $G_Y$ converts the speech data $x$ from speaker $X$ to that of speaker $Y$. The cycle consistency loss is defined as follows:

$$\mathcal{L}_{\text{cycle}}(G_X, G_Y) = \mathbb{E}_x\left[\left\|G_X(G_Y(x)) - x\right\|_1\right] + \mathbb{E}_y\left[\left\|G_Y(G_X(y)) - y\right\|_1\right]. \quad (2)$$

It means that the cyclic conversions should bring the final output data back to the original input data, which helps to keep the linguistic contents of the speech unchanged during conversion.

Since the generator should not modify the target speaker's voice when it is fed into the generator as the input, the identity mapping loss can be used additionally, which is defined as follows:

$$\mathcal{L}_{\text{identity}}(G_X, G_Y) = \mathbb{E}_x[\|G_X(x) - x\|_1] + \mathbb{E}_y[\|G_Y(y) - y\|_1] \, . \quad (3)$$

The objective function for the CycleGAN then becomes as follows:

$$\begin{aligned}\mathcal{L}_{\text{CycleGAN}}(G_X, D_X, G_Y, D_Y) = \\ \mathcal{L}_{\text{GAN}}(G_X, D_X) + \\ \mathcal{L}_{\text{GAN}}(G_Y, D_Y) + \\ \lambda_1 \mathcal{L}_{\text{cycle}}(G_X, G_Y) + \\ \lambda_2 \mathcal{L}_{\text{identity}}(G_X, G_Y) \, ,\end{aligned} \quad (4)$$

where $\lambda_1$ and $\lambda_2$ are the relative weights for the cycle consistency loss and the identity mapping loss, respectively.

The CycleGAN has been successfully applied to voice conversion and has shown the state-of-the-art results using non-parallel training data [10, 11]. However, it can handle only one-to-one voice conversion between two speakers. To overcome this limitation, we propose to use the CC-GAN for many-to-many voice conversion, which can simulate multiple CycleGANs using a single conditional GAN.

## 2.3. Conditional CycleGAN (CC-GAN)

In a CC-GAN, a pair of source and target speaker identity vectors is utilized to steer the GAN in synthesizing the speech for the target speaker [14]. As a result, the two GANs in a CycleGAN can be merged into one conditional GAN, and the generator of the resulting single conditional GAN is used to convert the voice for both directions, i.e., from speaker $X$ to speaker $Y$ and vice versa. This idea can be extended to many-to-many voice conversion for more than two speakers.

Figure 1 (a) describes the generator of the CC-GAN for many-to-many voice conversion. As in [10], it uses one-dimensional convolutional neural networks (CNN) with gated linear units (GLU) [15]. The generator consists of the first convolutional layer, two down-sampling layers, six residual blocks, two up-sampling layers, and the last convolutional layer. The first convolutional layer and the down-sampling layers take the source speaker identity vector and the output from the preceding layers as the input to the current layer. The convolutional layers in each residual block take both the source and the target speaker identity vectors as well as the output from the preceding layers. The up-sampling layer and the last convolutional layer take the target speaker identity vector and the output from the preceding layers.

Figure 1 (b) describes the multi-output discriminator of the CC-GAN for many-to-many voice conversion. It uses two-dimensional CNNs consisting of the first convolutional layer, three down-sampling layers, and a fully connected layer. Instead of having one discriminator for each speaker, the CC-GAN has a single discriminator that yields multiple outputs. Each output node of the multi-output discriminator is interpreted as the output of the corresponding speaker's discriminator. Therefore, it uses multiple sigmoid activation functions instead of a single softmax activation function. Each layer of the discriminator takes the target speaker identity vector and the output from the preceding layer as the input to the current layer.

Since the discriminator produces multiple output values, an additional step is needed to extract a single value for the loss computation. It can be done by a dot-product between the discriminator output and the target speaker identity vector that is represented as a one-hot vector, as follows:

$$D(x, X) \cdot X \, , \quad (5)$$

where '$\cdot$' represents the dot-product operator and $X$ denotes the speaker identity vector for utterance $x$. Since the position of value one in the one-hot vector indicates the corresponding speaker, the rest of the discriminator output values are simply discarded.

Thus, the GAN loss for the CC-GAN can be defined as follows:

$$\mathcal{L}'_{\text{GAN}}(G, D) = \mathbb{E}_{x, X|x}[\log D(x, X) \cdot X] + \mathbb{E}_{x, X|x, Y}[\log(1 - D(G(x, X, Y), Y) \cdot Y)] \, , (6)$$

where generator $G(x, X, Y)$ converts the speech data $x$ from speaker $X$ to that of speaker $Y$, and discriminator $D(x, X)$

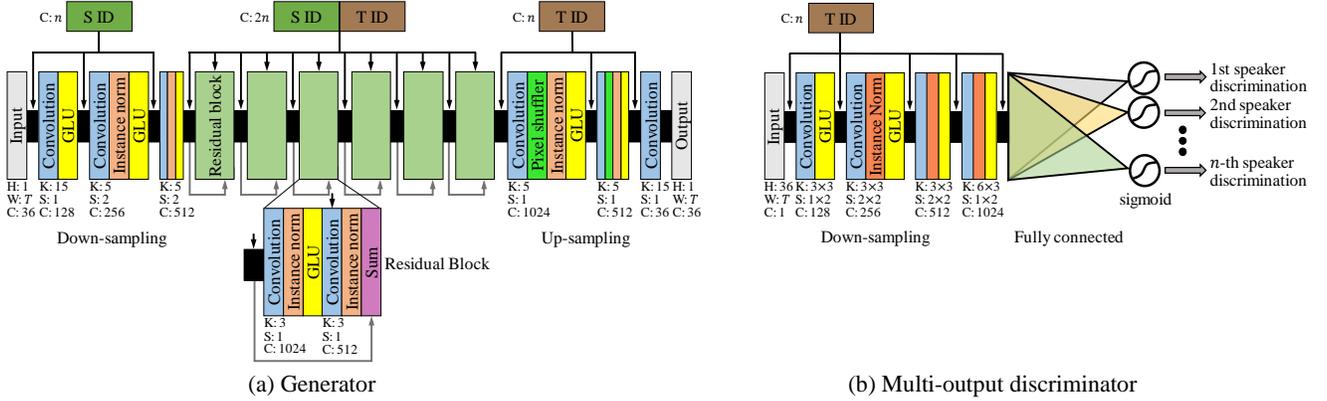

(a) Generator  (b) Multi-output discriminator

Fig. 1. The generator and the multi-output discriminator of the CC-GAN. 'S ID' and 'T ID' represent the source and the target speaker identity vectors, respectively. 'H', 'W', 'C', 'K', and 'S' represent height, width, number of channels, kernel size, and stride, respectively.

checks whether speech data $x$ belongs to speaker $X$. Both $X$ and $Y$ are represented as one-hot vectors.

Then, the cycle consistency loss for the CC-GAN is defined as follows:

$$\mathcal{L}'_{\text{cycle}}(G) = \mathbb{E}_{x,X|x,Y}[\|G(G(x,X,Y),Y,X) - x\|_1] . \quad (7)$$

Note that since there is only one generator in the CC-GAN, equation (7) is used to compute the cycle consistency loss for all pairs of speakers.

Now, the identity mapping loss for the CC-GAN is defined as follows:

$$\mathcal{L}'_{\text{identity}}(G) = \mathbb{E}_{x,X|x,Y}[\|G(x,Y,X) - x\|_1] . \quad (8)$$

Finally, the objective function for the CC-GAN can be written as follows:

$$\mathcal{L}_{\text{CC-GAN}}(G, D) = \mathcal{L}'_{\text{GAN}}(G, D) + \lambda_1 \mathcal{L}'_{\text{cycle}}(G) + \lambda_2 \mathcal{L}'_{\text{identity}}(G) . \quad (9)$$

We found experimentally that it is better to use the average loss from all output nodes of the discriminator instead of only the target speaker's output node when training the discriminator with the fake data. Also, we used the least square GAN loss [16] for model stabilization in the experiments.

The idea of many-to-many voice conversion by extending the CycleGAN first appeared in [14,18]. It was shown experimentally that the model size can be reduced by half for a two-speaker voice conversion case [14]. In the next section, we show how the CC-GAN for multi-speaker voice conversion performs when the number of speakers is more than two. Another extension of the CycleGAN for many-to-many voice conversion is the StarGAN [17] based voice conversion [18, 19]. It is similar to the CC-GAN based many-to-many voice conversion in that the GAN is conditioned on the speaker identity vectors. The difference is that the CC-GAN uses the source speaker identity vector, the target speaker identity vector, or both to condition all units in the GAN, making it fully conditional. Also, the discriminator and the domain classifier of the StarGAN is merged into one multi-output discriminator in the CC-GAN. To the best of our knowledge, these extensions of the CycleGAN for many-to-many voice conversion have never been compared experimentally with the multiple vanilla CycleGAN models for many-to-many voice conversion, which would be the performance upper bound of the extension approaches. In the next section, the performance of the proposed method is compared with the multiple CycleGANs as well as the StarGAN based approach.

## 3. EXPERIMENTS

We compared the proposed CC-GAN based method with two conventional approaches, i.e., the CycleGAN based VC [10] and the StarGAN based VC [18], using the VCC2018 corpus [20]. The speakers in the corpus consist of 6 males and 6 females. There are 116 parallel utterances for each speaker (81 utterances for training and 35 utterances for evaluation). 36 Mel-cepstral coefficients (MCCs), aperiodicities (APs), and logarithmic fundamental frequency (F0) were extracted every 5 ms from the speech waveforms of which sampling rate is 22.05 kHz. In all three methods, the generators of the GANs converted the MCCs only. The F0's were converted by the logarithm Gaussian normalized transformation [21], and the APs were used without any modification to synthesize the waveforms of the converted speech using the WORLD vocoder [22].

In every training iteration, we randomly paired the source and the target utterances and randomly extracted 128 MCC frames from each utterance to make the non-parallel training data condition. We used the Adam optimizer [23] with a batch size of 1. The learning rates of the generator and the discriminator were set to 0.0002 and 0.0001, respectively, and were linearly decreased after $1/2$ of the total number of epochs which was set to 950. The relative weights of $\mathcal{L}'_{\text{GAN}}$, $\mathcal{L}'_{\text{cycle}}$, and $\mathcal{L}'_{\text{identity}}$ were set to 4, 10, and 5 respectively.

While we were able to train a single CC-GAN for all 132 conversion directions among 12 speakers, we decided to train only 4 CycleGAN models for the 4 conversion directions used in [18] for the comparison as it was impractical to train all 66 CycleGAN models due to the time and memory constraints. On the other hand, since we were not able to reproduce the same results as in [18] for the StarGAN based VC, we downloaded the converted speech samples from the author's website and used them for the comparative evaluations. Though the number of sample utterances is small, we think this is fair comparison. We also trained a 4-speaker CC-GAN model for the fair comparison with [18].

Table 1. MCD with standard deviation

|  | CycleGAN | StarGAN | CC-GAN |
|---|---|---|---|
| F to F | 6.47 ± 0.22 | 8.07 ± 0.34 | 7.25 ± 0.24 |
| M to F | 7.24 ± 0.15 | 9.30 ± 0.54 | 7.85 ± 0.14 |
| F to M | 6.79 ± 0.20 | 8.57 ± 0.10 | 7.30 ± 0.38 |
| M to M | 6.78 ± 0.23 | 8.35 ± 0.06 | 7.45 ± 0.10 |
| Average | 6.82 ± 0.34 | 8.57 ± 0.56 | 7.46 ± 0.34 |

Table 2. MSD with standard deviation

|  | CycleGAN | StarGAN | CC-GAN |
|---|---|---|---|
| F to F | 1.87 ± 0.04 | 1.87 ± 0.03 | 1.86 ± 0.04 |
| M to F | 1.84 ± 0.01 | 1.86 ± 0.02 | 1.83 ± 0.01 |
| F to M | 1.83 ± 0.01 | 1.82 ± 0.01 | 1.85 ± 0.00 |
| M to M | 2.00 ± 0.26 | 1.86 ± 0.02 | 1.89 ± 0.08 |
| Average | 1.88 ± 0.15 | 1.85 ± 0.03 | 1.86 ± 0.05 |

Table 3. Sound quality test (MOS and standard deviation)

|  | StarGAN | CC-GAN | Target Voice |
|---|---|---|---|
| F to F | 4.03 ± 0.83 | 3.61 ± 0.86 | 4.81 ± 0.49 |
| M to F | 2.22 ± 0.75 | 3.00 ± 1.05 | 4.81 ± 0.49 |
| F to M | 3.39 ± 0.86 | 3.92 ± 0.98 | 4.79 ± 0.41 |
| M to M | 3.53 ± 1.04 | 3.81 ± 0.81 | 4.79 ± 0.41 |
| Average | 3.29 ± 1.10 | 3.58 ± 1.00 | 4.80 ± 0.45 |

Table 4. Similarity test (%)

|  | StarGAN | Fair | CC-GAN |
|---|---|---|---|
| F to F | 22% | 50% | 28% |
| M to F | 25% | 6% | 69% |
| F to M | 8% | 42% | 50% |
| M to M | 17% | 39% | 44% |
| Average | 18% | 34% | 48% |

We conducted objective evaluations as well as subjective evaluations. For the objective evaluations, we used two measures, i.e., the Mel-cepstral distortion (MCD) [2] and the modulation spectral distance (MSD) [24] between the real speech from the target speaker and the converted speech. Tables 1 and 2 show the MCD and the MSD, respectively, computed using 12 utterances (3 evaluation utterances for each of the target speakers SF1, SM1, SF2, and SM2). The CC-GAN based VC outperforms the StarGAN based VC in terms of the MCD.

For the subjective evaluations, we conducted sound quality tests and similarity tests. The mean opinion score (MOS) was used for the sound quality test, where the speech utterances from target speakers and the two conversion methods were played one at a time in random order and the 12 listeners were asked to evaluate the naturalness of each speech by selecting scores between 1 and 5 (1 being bad and 5 being good). In the similarity test, a real speech utterance from a target speaker was played first, and each of the converted speech from the two conversion methods were played in random order. The 12 listeners were asked to select the most similar utterance to the target speaker's speech or 'fair' if they cannot tell the difference.

Table 3 shows the results of the sound quality test. The CC-GAN based VC shows better performance than the StarGAN based VC. Table 4 shows the results of the similarity test. The CC-GAN based VC is ahead of the StarGAN based VC in the similarity test.

## 4. CONCLUSIONS

We proposed a novel many-to-many non-parallel voice conversion method called the CC-GAN based VC. It uses only a single GAN for many-to-many voice conversion while the CycleGAN based VC would require $n(n-1)$ GANs for $n$ speakers. As a result, the CC-GAN based VC decreases the training time significantly as well as the model size for many-to-many voice conversion. We showed experimentally that the proposed method was comparable to the CycleGAN based VC which shows the state-of-the-art performance for one-to-one non-parallel voice conversion without using any extra modules such as ASR systems. To the best of our knowledge it is the first work that shows the feasibility of the extension of the CycleGAN for many-to-many voice conversion using 12 speakers.

As for the future research, we plan to build a CC-GAN with hundreds of speakers for the voice conversion of unseen speakers by using i-vectors [25] or x-vectors [26] to condition the GAN. Since CycleGAN-VC2 [11] and StarGAN-VC2 [19] were published after we developed the CycleGAN and the CC-GAN, we did not have enough time to include the additional features of [11] (i.e., two-step adversarial loss, 2-1-2D CNN, and PatchGAN) into our implementations of CycleGAN and CC-GAN. Including these new features and replacing the vocoder with powerful neural vocoders such as WaveNet [27] or WaveRNN [28] can be another future research direction.

## 5. ACKNOWLEDGEMENTS

This research was supported by the Basic Science Research Program through the National Research Foundation of Korea (NRF) funded by the Ministry of Science, ICT and Future Planning (NRF-2017R1E1A1A01078157). Also, it was partly supported by the MSIT (Ministry of Science and ICT) under the ITRC (Information Technology Research Center) support program (IITP-2018-0-01405) supervised by the IITP (Institute for Information & Communications Technology Planning & Evaluation), and IITP grant funded by the Korean government (MSIT) (No. 2018-0-00269).

## 6. REFERENCES

[1] Y. Stylianou, O. Cappe, and E. Moulines, "Continuous probabilistic transform for voice conversion," *IEEE Transactions on Speech and Audio Processing*, vol. 6, no. 2, pp. 131–142, 1998.

[2] T. Toda, A. Black, and K. Tokuda, "Voice conversion based on maximum-likelihood estimation of spectral parameter trajectory," *IEEE Transactions on Audio, Speech, and Language Processing*, vol. 15, no. 8, pp. 2222–2235, 2007.

[3] E, Helander, T. Virtanen, J. Nurminen, and M. Gabbouj, "Voice conversion using partial least squares regression," *IEEE Transactions on Audio, Speech, and Language Processing*, vol. 18, no. 5, pp. 912–921, 2010.

[4] C. Hsu, H. Hwang, Y. Wu, Y. Tsao, and H. Wang, "Voice conversion from non-parallel corpora using variational auto-encoder," *Asia-Pacific Signal and Information Processing Association Annual Summit and Conference*, 2016.

[5] C. Hsu, H. Hwang, Y. Wu, Y. Tsao, and H. Wang, "Voice conversion from unaligned corpora using variational autoencoding Wasserstein generative adversarial networks," *Interspeech*, pp. 3364–3368, 2017.

[6] A. Oord, O. Vinyals, and K. Kavukcuoglu, "Neural discrete representation learning," *Conference on Neural Information Processing Systems*, pp. 6306–6315, 2017.

[7] J. Chou, C. Yeh, H. Lee, and L. Lee, "Multi-target voice conversion without parallel data by adversarially learning disentangled audio representations," *Interspeech*, pp. 501–505, 2018.

[8] S. Liu, J. Zhong, L. Sun, X. Wu, X. Liu and H. Meng, "Voice conversion across arbitrary speakers based on a single target-speaker utterance," *Interspeech*, pp. 496–500, 2018.

[9] H. Kameoka, T. Kaneko, K. Tanaka, and N. Hojo, "ACVAE-VC: Non-parallel voice conversion with auxiliary classifier variational autoencoder," *IEEE/ACM Transactions on Audio, Speech, and Language Processing*, vol. 27, no. 9, pp. 1432–1443, 2019.

[10] T. Kaneko and H. Kameoka, "CycleGAN-VC: Non-parallel voice conversion using cycle-consistent adversarial networks," *European Signal Processing Conference*, pp. 2114–2118, 2018.

[11] T. Kaneko, H. Kameoka, K. Tanaka, and N. Hojo, "CycleGAN-VC2: Improved CycleGAN-based non-parallel voice conversion," *IEEE International Conference on Acoustics, Speech, and Signal Processing*, pp. 6820–6824, 2019.

[12] J. Zhu, T. Park, P. Isola, and A. Efros, "Unpaired image-to-image translation using cycle-consistent adversarial networks," *IEEE International Conference on Computer Vision*, pp. 2242–2251, 2017.

[13] I. Goodfellow, J. Pouget-Abadie, M. Mirza, B. Xu, D. Warde-Farley, S. Ozair, A. Courville, Y. Bengio, "Generative adversarial nets," *Conference on Neural Information Processing Systems*, pp. 2672–2680, 2014.

[14] D. Yook, I. Yoo, and S. Yoo, "Voice conversion using conditional CycleGAN," *International Conference on Computational Science and Computational Intelligence*, pp. 1460–1461, 2018.

[15] Y. Dauphin, A. Fan, M, Auli, and D. Grangier, "Language modeling with gated convolutional networks," *International Conference on Machine Learning*, pp. 933–941, 2017.

[16] X. Mao, Q. Li, H. Xie, R. Lau, Z. Wang, and S. Smolley, "Least squares generative adversarial networks," *IEEE International Conference on Computer Vision*, pp. 2794–2802, 2017.

[17] Y. Choi, M. Choi, M. Kim, J. Ha, S. Kim, and J. Choo, "StarGAN: Unified generative adversarial networks for multi-domain image-to-image translation," *IEEE Conference on Computer Vision and Pattern Recognition*, pp. 8789–8797, 2018.

[18] H. Kameoka, T. Kaneko, K. Tanaka, and N. Hojo, "StarGAN-VC: Non-parallel many-to-many voice conversion using star generative adversarial networks," *IEEE Workshop on Spoken Language Technology*, pp. 266–273, 2018.

[19] T. Kaneko, H. Kameoka, K. Tanaka, and N. Hojo, "StarGAN-VC2: Rethinking conditional methods for StarGAN-based voice conversion," *Interspeech*, pp. 679–683, 2019.

[20] J. Lorenzo-Trueba1, J. Yamagishi, T. Toda, D. Saito, F. Villavicencio, T. Kinnunen, and Z. Ling, "The voice conversion challenge 2018: Promoting development of parallel and nonparallel methods," *The Speaker and Language Recognition Workshop*, pp. 195–202, 2018.

[21] K. Liu, J. Zhang, and Y. Yan, "High quality voice conversion through phoneme-based linear mapping functions with STRAIGHT for Mandarin," *International Conference on Fuzzy Systems and Knowledge Discovery*, 2017.

[22] M. Morise, F. Yokomori, and K. Ozawa, "WORLD: A vocoder-based high-quality speech synthesis system for real-time applications," *IEICE Transactions on Information and Systems*, vol. E99.D, no. 7, pp. 1877–1884, 2016.

[23] D. Kingma and J. Ba, "Adam: A method for stochastic optimization," *International Conference on Learning Representations*, 2015.

[24] S. Takamichi, T. Toda, A. Black, G. Neubig, S. Sakti, and S. Nakamura, "Postfilters to modify the modulation spectrum for statistical parametric speech synthesis," *IEEE/ACM Transactions on Audio, Speech, and Language Processing*, vol. 24, no. 4, pp. 755–767, 2016.

[25] N. Dehak, P. Kenny, R. Dehak, P. Dumouchel, and P. Ouellet, "Front-end factor analysis for speaker verification," *IEEE Transactions on Audio, Speech, and Language Processing*, vol. 19, no. 4, pp. 788–798, 2011.

[26] D. Snyder, D. Garcia-Romero, G. Sell, D. Povey, and S. Khudanpur, "X-Vectors: Robust DNN embeddings for speaker recognition," *IEEE International Conference on Acoustics, Speech and Signal Processing*, pp. 5329–5333, 2018.

[27] A. Oord, S. Dieleman, H. Zen, K. Simonyan, O. Vinyals, A. Graves, N. Kalchbrenner, A. Senior, and K. Kavukcuoglu, "WaveNet: A generative model for raw audio," arXiv:1609.03499, 2016.

[28] N. Kalchbrenner, E. Elsen, K. Simonyan, S. Noury, N. Casagrande, E. Lockhart, F. Stimberg, A. Oord, S. Dieleman, and K. Kavukcuoglu, "Efficient neural audio synthesis," arXiv:1802.08435, 2018.